\documentclass[11pt,twoside]{article}
\usepackage{FUSE2004}
\usepackage{natbib}

\usepackage{epsf}
\usepackage{psfig}
\usepackage{lscape}

\begin{document}
\title{Insight into AGB and post-AGB stellar evolution with FUSE}
\author{K. Werner, T. Rauch}
\affil{Institut f\"ur Astronomie und Astrophysik, Univ.\ T\"ubingen, Germany}
\author{J.W. Kruk}
\affil{Department of Physics and Astronomy, JHU, Baltimore MD, U.S.A.}

\begin{abstract}
FUSE spectroscopy has proved that extremely hot hydrogen-deficient post-AGB
stars (PG1159 stars) display matter on their surface that usually remains hidden
in the region between the H- and He-burning shells of the former AGB
star. Hence, the spectral analysis of PG1159 stars allows to study directly the
chemistry of this intershell region which is the outcome of complicated burning
and mixing processes during AGB evolution. Detailed abundance determinations
provide constraints for these processes which are still poorly understood. With
FUSE we have discovered high neon and fluorine overabundances.  There is also a
significant iron deficiency, which may be caused by  s-process neutron capture
transforming iron into heavier elements.
\end{abstract}

\section{Introduction}

PG1159 stars are hot hydrogen-deficient (pre-) white dwarfs ($T_{\rm eff}$
between 75\,000 and 200\,000\,K, $\log g$=5.5--8; Werner 2001). They are
probably the outcome of a late He-shell flash, a phenomenon that drives the
currently observed fast evolutionary rates of three well-known objects (FG~Sge,
Sakurai's object, V605 Aql). Flash-induced envelope mixing produces a
H-deficient stellar surface (Herwig et\,al.\ 1999). The photospheric composition
then essentially reflects that of the region between the H- and He-burning
shells in the precursor AGB star. The He-shell flash transforms the star back to
an AGB star (``born-again AGB star'') and the subsequent, second post-AGB
evolution explains the existence of Wolf-Rayet central stars of planetary
nebulae and their successors, the PG1159 stars.

PG1159 stars provide the unique possibility to study the chemistry in the
intershell region between the H- and He-burning shells that is created after
complicated and still poorly-understood burning and mixing processes during the
AGB phase. Usually the intershell material remains hidden within the stellar
interior. During the third dredge-up on the AGB, however, intershell material
can get mixed into the convective surface layer and appears on the stellar
surface, though in rather diluted abundances. Nevertheless, this process defines
the role of AGB stars as contributors of nuclearly processed matter to the
Galaxy. Our motivation to study PG1159 stars is based on the fact that these
objects directly display their intershell matter. However, the quantitative
interpretation of the abundance analyses is still premature because evolutionary
calculations through a final He-shell flash including a full nuclear network are
not available, yet.

Before the advent of FUSE with its outstanding FUV capabilities we have
performed spectral analyses of almost all known PG1159 stars based on optical
and HST-UV spectra. We determined $T_{\rm eff}$, $\log g$, and abundances of the
dominant elements, namely He, C, and O. Typical values are He=33\%, C=50\%,
O=17\% (by mass). Traces of nitrogen (1\%) or considerable amounts of residual
hydrogen (about 25\%) were found in a few stars, and in a few cases, optical as
well as Chandra X-ray identifications of neon were successful (2\%).

\begin{figure}
\epsfxsize=\textwidth
\epsffile{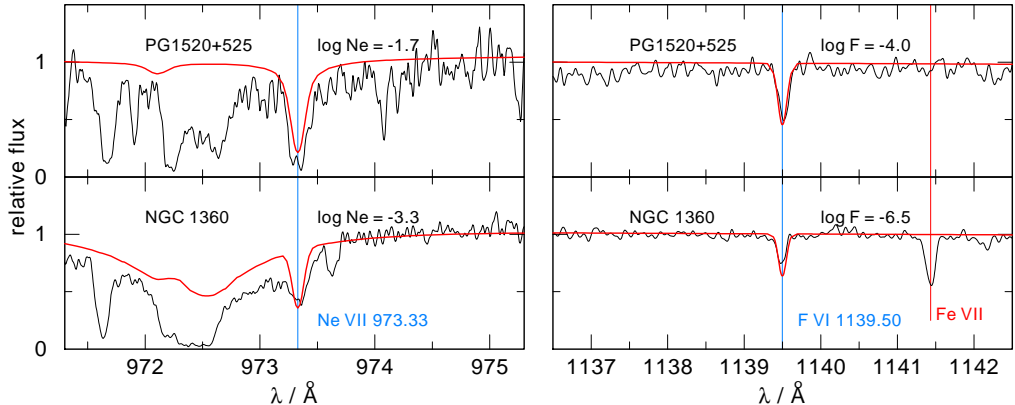}
\caption{
Discovery of a neon line (left panels) and a fluorine line (right panels) in the
hydrogen-deficient PG1159-type central star PG\,1520+525 (top panels) and in the
hydrogen-normal central star of NGC\,1360 (bottom panels). The neon and fluorine
abundances in the PG1159 star (given as mass fractions in the panels) are
strongly enhanced, namely 20 times and 250 times solar, respectively, whereas
they are solar in NGC\,1360. Note the strong Fe\,VII line (not included in the
models) at 1141.4\AA\ in NGC\,1360, which indicates a solar iron abundance
(Hoffmann et\,al.\ 2005). It is not detectable in the PG1159 star, probably due
to a subsolar Fe abundance.}
\end{figure}

\section{Most important results from FUSE spectroscopy}

Generally, FUSE spectra of PG1159 stars show only few photospheric (absorption)
lines, mainly from He\,II, C\,IV, O\,VI, and Ne\,VII. Some of them show shallow
N\,V lines and in many of them we see sulfur. The S\,VI 933/944\AA\ doublet in
K1-16 suggests a solar abundance, which is in line with the expectation that S
is not affected by the s-process. We also identify silicon in some objects
(Reiff et\,al.\ 2005), but detailed abundance analyses remain to be done.

The first surprising result of FUSE spectroscopy was the detection of a
significant iron deficiency (1--2 dex) in the three best studied PG1159 stars
(Miksa et\,al.\ 2002).  Apparently, Fe was transformed to heavier elements in
the intershell region of the AGB star  by n-captures from the neutron source
$^{13}$C($\alpha$,n)$^{16}$O (Herwig et\,al.\ 2003). Subsequently, several other
studies have also revealed an Fe-deficiency in [WC] type central stars (see
Werner et\,al.\ 2003), which matches our picture that these stars are immediate
PG1159 star progenitors.

The next important result was accomplished by the identification of one of the
strongest absorption lines seen in FUSE spectra of most PG1159 stars, located at
973.3\AA. It is a Ne\,VII line (Fig.\,1) that allowed to assess the neon
abundance in a large sample of objects (Werner et\,al.\ 2004). It turns out that
neon is strongly overabundant, (2\%, i.e., 20 times solar). This result clearly
confirms the idea that PG1159 stars indeed exhibit intershell matter. Neon is
produced in the He-burning environment by two $\alpha$-captures  of nitrogen,
which itself resulted from previous CNO burning:
$^{14}$N$(\alpha,\gamma)^{18}$F$({\rm e}^+\nu)^{18}$O$(\alpha,\gamma)^{22}$Ne.

There are still at least ten photospheric lines in the FUSE spectra of PG1159
stars that remain unidentified. Some of them may stem from yet unknown Ne\,VII
lines. The latest identification is a feature at 1139.5\AA, which appears rather
strong in some objects. We found that it is a line from highly ionized fluorine
(Fig.\,1) and derived large overabundances (up to 250 times solar) for a number
of PG1159 stars. We also could identify this line in ``normal'' H-dominated
central stars and, in contrast, find about solar fluorine abundances (Werner
et\,al.\ 2005). This again is a clear proof that we see intershell matter on
PG1159 stars. According to recent calculations by Lugaro et\,al.\ (2004), their
stellar models show an effective fluorine production and storage in the
intershell, leading to abundances that are comparable to the observed PG1159
abundances of fluorine.  The general problem for fluorine production is that
$^{19}$F, the only stable F isotope, is rather fragile and readily destroyed in
hot stellar interiors by hydrogen via $^{19}$F(p,$\alpha$)$^{16}$O and helium
via $^{19}$F($\alpha$,p)$^{22}$Ne. The nucleosynthesis path for F production in
He-burning environments of AGB and Wolf-Rayet stars is
$^{14}$N($\alpha$,$\gamma$)$^{18}$F($\beta^+$)$^{18}$O(p,$\alpha$)$^{15}$N($\alpha$,$\gamma$)$^{19}$F.

This underlines that AGB stars which dredge up material from the intershell are
contributing to the Galactic F content (together with Wolf-Rayet stars and
type~II SNe). This is completely in line with the detected F overabundances (up
to 30 times solar) found in AGB stars (Jorissen et\,al.\ 1992). To what extent
PG1159 stars themselves return F to the ISM remains to be estimated. The life
time of a born-again AGB star is short in comparison to a usual AGB star,
however, the F fraction in the mass lost by a wind of the former is much higher.

\acknowledgements This work is supported by DLR (50\,OR\,0201), DFG 
(WE\,1312/30-1), and the FUSE project, funded by NASA contract NAS5-32985.

\end{document}